\documentclass[a4paper,12pt]{article}

\usepackage[utf8]{inputenc} 

\usepackage{amsfonts}
\usepackage{authblk}
\usepackage{amssymb}
\usepackage{amsbsy}
\usepackage[margin=2cm]{geometry}

\usepackage{xcolor}
\usepackage{cite}
\usepackage{mathrsfs}
\usepackage{hyperref}
\usepackage{graphicx} 

\usepackage{booktabs} 
\usepackage{array} 
\usepackage{paralist} 
\usepackage{verbatim} 

\usepackage{subfigure}
\makeatletter
\renewcommand{\@thesubfigure}{\normalsize(\textbf{\alph{subfigure}})}
\makeatother

\usepackage{fancyhdr} 
\title{A New Mechanism of Open System Evolution and Its Entropy Using Unitary Transformations in Noncomposite Qudit Systems}

\author[1,2]{Julio A. L\'opez-Sald\'ivar}
\author[1]{Octavio Casta\~nos}
\author[3]{Margarita A. Man'ko}
\author[2,3,4]{Vladimir I. Man'ko}
\affil[1]{Instituto de Ciencias Nucleares, Universidad Nacional Aut\'onoma de M\'exico,
Apdo. Postal 70-543, Ciudad de México 04510, M\'exico}
\affil[2]{Moscow Institute of Physics and Technology (State University), Institutskii per. 9,
Dolgoprudnyi, Moscow~Region 141700, Russia}
\affil[3]{Lebedev Physical Institute, Russian Academy of Sciences, Leninskii Prospect 53,
Moscow 119991, Russia}
\affil[4]{Department of Physics, Tomsk State University, Lenin Avenue 36, Tomsk 634050, Russia}
\date{}

\begin{document}

\maketitle

\begin{abstract}
The evolution of an open system is usually associated with the interaction of
the system with an environment. A new method to study the open-type
system evolution of a qubit \mbox{(two-level atom)} state is established.~This evolution is determined by a unitary transformation applied to 
the qutrit (three-level atom) state, which defines the qubit subsystems.~This procedure can be used to obtain different qubit quantum channels
employing unitary transformations into the qutrit system. In~particular, we
study the phase damping and spontaneous-emission quantum channels.
In addition, we mention a proposal for quasiunitary transforms of
qubits, in view of the unitary transform of the total qutrit system. The
experimental realization is also addressed. The probability
representation of the evolution and its information-entropic
characteristics are considered.
\end{abstract}

\section{Introduction}

The open system evolution of a qudit state is known to be the result
of interactions with an environment.~Usually, the~states of the
complete system are thought to evolve by a unitary transformation in the
Hilbert space $\hat{H}=\hat{H}_q \otimes \hat{H}_{env}$, then the
density operator of the composite system leads us, using the partial
tracing procedure, to~the density operator of the subsystem
$\hat{\rho}_q$ (qudit), and~its evolution is induced by the unitary
evolution of the complete system. In~this picture, the~qubit state
dynamics needs the structure of the Hilbert space $\hat{H}$
corresponding to the presence of two subsystems, qudit and
environment~\cite{weiss}. In~this work, we suggest a new mechanism to study the
open system evolution, which does not demand the complete system to
have a~subsystem.

We show that for any system without subsystems, there exist a unitary
evolution, which due to hidden correlations in the system,
evolves according to the Gorini--Kossakowski--Sudarshan--Lindblad
equation~\cite{kossa,ingarden,gorini,lindblad}. We demonstrate this picture using the example of a qutrit
(complete system without subsystems), where the open-like evolution
is available for their associated~qubits.

In previous
works~\cite{chernegaA,chernegaB,chernegaC,entropy18,QIP}, a~new
method to define different qubit density matrices from a qudit
system was established. This procedure uses the occupation
probabilities and transition probability amplitudes for different
levels of a qudit system and groups them as if there exists two levels only. This~is done by mapping the qudit density matrix to the closest higher
even-dimensional density matrix. The~partial trace operation then is enacted on the
resulting matrix in order to obtain well-defined qubit density
matrices.

The obtained qubits have been used to define a new geometric
representation of the $d$-dimensional qudit states through $d$ Bloch
vectors~\cite{QIP} associated with the generated qubits. Furthermore, it has been 
possible to describe quantum phenomena as the entanglement on a
two-qubit system in terms of standard
probabilities~\cite{entropy18}.

The evolution of a qutrit density matrix can provide the quantum
channel, which maps the initial state $\hat{\rho}_a$ onto the
density matrix $\hat{\rho}_a^\prime$. The~proposed open-type evolution establishes a new mechanism, which will need a special state preparation and a specific unitary operation for the
qutrit system, as~we will show later on. The~experimental possibilities by which one can
realize this new mechanism are related to superconducting
circuit devices~\cite{wallraff,superc}.

Most quantum computing processes consider a set of pure qubit states,
which are transformed by unitary operators, also called gates, that
are used to implement different computing algorithms. In~this
article, instead, we have density matrices (which might be describing a mixed state) of
larger qudit systems. The~definition of a set of qubit states from a
qudit system is similar to the ideas established in~\cite{neeley}, where the
emulation of a spin system was obtained from qudit states, and~in~\cite{lanyon}, where the quantum logic of qubits was simplified by
the use of a higher dimensional Hilbert space; and in general, with~all the
procedures that make use of larger Hilbert spaces. In~this work, we
demonstrate that subsystems of qubits defined by larger systems can be used in
quantum information. A~principal foundation of quantum computation
is the study of quantum channels. These channels are linked to
unitary transformations of the qubit density matrix. There exist several
channels that can describe the interaction between a quantum
system and its environment such as the bit-flip, depolarization,
spontaneous emission, phase, and~amplitude damping channels. For~this, the~study of quantum channels has been of relevance in the
error correction theory of quantum computation~\cite{chuang,terhal}.

Here, we present different examples of quantum channels, which act
on the associated qubits to qudit states. These quantum channels
have the advantage of being represented as unitary transformations acting in
the qudit system, providing the possibility to study the qubits as
if they were interacting with an~environment.

On the other hand, the~study of the interaction of three-level
systems with electromagnetic fields has led to the discovery of
important phenomena, such as the presence of dark states~\cite{dark1976}
together with black resonances~\cite{reso1987} and~electromagnetically-induced
transparency~\cite{eit1990,eit1993,eit2005}. This is important to
our objectives as in some cases, the herein proposed qubit quantum channels can be obtained by a unitary transformation of dark states,
suggesting the possibility of checking our results~experimentally.

The work is organized as follows: In Section~\ref{sec2}, a~review of the qubit density matrices that are associated with a qutrit state is given.~Furthermore, the~association of a unitary transform of the qutrit to the nonunitary transformations of the qubits is studied.~In~Section~\ref{sec3}, the~definitions of the qubit phase damping and spontaneous-emission quantum channels are reviewed.~Later, the~unitary transformations of a qutrit system are explicitly given, which yields the phase damping and spontaneous-emission channels on the associated qubits.~A~way to obtain a quasi-unitary transformation on the qubits is also explored. The~change of entropy associated with the nonunitary evolution of the qubits is discussed in Section~\ref{sec4}. Finally, some concluding remarks are~given. 

\section{Nonunitary Evolution for the Qubit Decomposition of Qutrit~States}\label{sec2}
In a previous work~\cite{QIP}, we showed the existence of six
different qubit states associated with a general qutrit density matrix:
\[
\hat{\rho}=\left( \begin{array}{ccc}
\rho_{11} & \rho_{12} & \rho_{13} \\
\rho_{21} & \rho_{22} & \rho_{23} \\
\rho_{31} & \rho_{32} & \rho_{33}
\end{array}\right) \, .
\]

To define these states, different maps of $\hat{\rho}$ to a
4 $\times$ 4 density matrix, with~one row and one column equal to zero
(in such a way that ensures an eigenvalue equal to zero), were used.
Then, the~partial trace of the resulting 4 $\times$ 4 matrix was
performed as if it was describing a two-qubit system. The~obtained qubit
partial density operators can be explicitly written as:
\begin{eqnarray}
\hat{\rho}_1=\left( \begin{array}{cc}1-\rho_{33} & \rho_{13} \\ \rho_{31} & \rho_{33} \end{array} \right) , \quad
\hat{\rho}_2=\left( \begin{array}{cc}1-\rho_{22} & \rho_{12} \\ \rho_{21} & \rho_{22} \end{array} \right) , \quad
\hat{\rho}_3=\left( \begin{array}{cc}\rho_{11} & \rho_{13} \\ \rho_{31} & 1-\rho_{11} \end{array} \right) , \nonumber \\
\hat{\rho}_4=\left( \begin{array}{cc}\rho_{22} & \rho_{23} \\ \rho_{32} & 1-\rho_{22} \end{array} \right) , \quad
\hat{\rho}_5=\left( \begin{array}{cc}\rho_{11} & \rho_{12} \\ \rho_{21} & 1-\rho_{11} \end{array} \right) , \quad
\hat{\rho}_6=\left( \begin{array}{cc}1-\rho_{33} & \rho_{23} \\ \rho_{32} & \rho_{33} \end{array} \right) .
\label{qubits}
\end{eqnarray}

The qubit states can be characterized in different sets by their corresponding von Neumann entropy
$S_k=-\mbox{Tr}\,\rho_k\ln\rho_k$, with~$k=1,2,\ldots,6$.~These qubits
correspond to the reduction of the three-level system to different
two-level systems by the summation of the population probabilities of
two levels into~one.

When the qutrit state is transformed using a general three-dimensional unitary matrix
$\hat{U}$, i.e.,~\mbox{$\hat{\rho}'=\hat{U}^\dagger\, \hat{\rho} \,
\hat{U}$}, the~qubits in Equation~(\ref{qubits}) are transformed in
a nonunitary way. The~transformed qubit density matrices can be written by
the following expressions:
\begin{eqnarray}
\hat{\rho}_1^\prime&=&\frac{1}{D}\left(\begin{array}{cc}D-M_{3,1} N_{1,3}+M_{2,1} N_{2,3}-M_{1,1} N_{3,3} &
M_{3,3} N_{1,3}-M_{2,3} N_{2,3}+M_{1,3} N_{3,3} \\ M_{3,1} N_{1,1}-M_{2,1} N_{2,1}+M_{1,1} N_{3,1} &
M_{3,1} N_{1,3}-M_{2,1} N_{2,3}+M_{1,1} N_{3,3}\end{array}\right), \nonumber \\
\hat{\rho}_2^\prime&=&\frac{1}{D}\left( \begin{array}{cc}D+ M_{3,2} N_{1,2}-M_{2,2} N_{2,2}+M_{1,2} N_{3,2} &
M_{3,3} N_{1,2}-M_{2,3} N_{2,2}+M_{1,3} N_{3,2} \\ -M_{3,2} N_{1,1}+M_{2,2} N_{2,1}-M_{1,2} N_{3,1} &
-M_{3,2} N_{1,2}+M_{2,2} N_{2,2}-M_{1,2} N_{3,2} \end{array}\right) , \nonumber \\
\hat{\rho}_3^\prime&=&\frac{1}{D}\left( \begin{array}{cc} M_{3,3} N_{1,1}-M_{2,3} N_{2,1}+M_{1,3} N_{3,1} &
M_{3,3} N_{1,3}-M_{2,3} N_{2,3}+M_{1,3} N_{3,3} \\ M_{3,1} N_{1,1}-M_{2,1} N_{2,1}+M_{1,1} N_{3,1} &
D-M_{3,3} N_{1,1}+M_{2,3} N_{2,1}-M_{1,3} N_{3,1}\end{array}\right) , \nonumber \\
\hat{\rho}_4^\prime&=&\frac{1}{D}\left(\begin{array}{cc}-M_{3,2} N_{1,2}+M_{2,2} N_{2,2}-M_{1,2} N_{3,2} &
-M_{3,2} N_{1,3}+M_{2,2} N_{2,3}-M_{1,2} N_{3,3} \\ M_{3,1} N_{1,2}-M_{2,1} N_{2,2}+M_{1,1} N_{3,2} &
D+M_{3,2} N_{1,2}-M_{2,2} N_{2,2}+M_{1,2} N_{3,2} \end{array} \right), \nonumber \\
\hat{\rho}_5^\prime&=&\frac{1}{D} \left(\begin{array}{cc} M_{3,3} N_{1,1}-M_{2,3} N_{2,1}+M_{1,3} N_{3,1} &
M_{3,3} N_{1,2}-M_{2,3} N_{2,2}+M_{1,3} N_{3,2} \\ -M_{3,2} N_{1,1}+M_{2,2} N_{2,1}-M_{1,2} N_{3,1} &
D-M_{3,3} N_{1,1}+M_{2,3} N_{2,1}-M_{1,3} N_{3,1}\end{array}\right) , \nonumber \\
\hat{\rho}_6^\prime&=&\frac{1}{D}\left(\begin{array}{cc} D-M_{3,1} N_{1,3}+M_{2,1} N_{2,3}-M_{1,1} N_{3,3} &
-M_{3,2} N_{1,3}+M_{2,2} N_{2,3}-M_{1,2} N_{3,3} \\ M_{3,1} N_{1,2}-M_{2,1} N_{2,2}+M_{1,1} N_{3,2} &
M_{3,1} N_{1,3}-M_{2,1} N_{2,3}+M_{1,1} N_{3,3}\end{array}\right) ,
\label{primes}
\end{eqnarray}
where $N_{jk}=(\hat{\rho} \hat{U})_{jk}$, $D$ is the determinant of $\hat{U}$, and~$M_{jk}$ are the
components of the minors of matrix $\hat{U}$, i.e.,~its elements
are the determinants after eliminating the $(4-j)^{\rm{th}}$ row and
$(4-k)^{\rm{th}}$ column of $\hat{U}$. The~transformed states are
characterized into different sets by their corresponding transformed entropies
$S'_k=-\mbox{Tr}\,\rho'_k\ln\rho'_k$. We emphasize that the
resulting qubit density matrices are associated, in~general, with~a nonunitary
evolution of the original qubits. This fact establishes a new
mechanism to obtain the open-like system evolution in a noncomposite
qutrit system. Additionally, this procedure can be extended to any qudit system,
in view of the general definition of the qubit density matrices obtained from a
qudit system~\cite{QIP}.

In~\cite{entropy18}, we discussed that a two-qubit density matrix
with one of its rows and columns equal to zero describes separable
states, if~one of the off-diagonal terms is equal to zero, for~example, the~state:
\[
\hat{\rho}=\left(
\begin{array}{cccc}
\rho_{11} & \rho_{12} & \rho_{13} & 0 \\
\rho_{21} & \rho_{22} & \rho_{23} & 0 \\
\rho_{31} & \rho_{32} & \rho_{33} & 0 \\
0 & 0 & 0 & 0
\end{array}
 \right)
\]
is separable iff $\rho_{23}=0$. {To show this, one can consider the previous density matrix to be in the standard two-qubit representation $\vert 00 \rangle$, $\vert 01 \rangle$, $\vert 10 \rangle$, and~$\vert 11 \rangle$. It can be seen that the partial transpose operation~\cite{horo} implies the change $\rho_{12}\leftrightarrow \rho_{21}$, and~for this reason, the eigenvalues of $\hat{\rho}$ with $\rho_{23}=0$ are equal to the eigenvalues of its partial transpose. As~the partial transpose is a nonnegative operator, then the system is separable}. The~separability
implies the invariance of the partial density matrices under local
unitary transformations. As~this two-qubit density matrix has a pair of
row-column with a diagonal term equal to zero, the~correspondence with a qutrit density
matrix can be made. On~the other hand, the~correspondence between
two-qubit local unitary transformations and qutrit unitary transformations can
be made in the same way, e.g.,~by eliminating one row and one column
of the two-qubit local transformation. This procedure allows us to define
different unitary transformations that almost leave the qubits
in Expression~(\ref{qubits}) invariant.

\section{Phase Damping and Spontaneous-Emission~Channels.}\label{sec3}
It is known that the interaction of a qubit system with an
environment leads to several physical phenomena such as dissipation
and decoherence in the qubit subsystem; an example of these
interactions is the phase damping channel. In~this channel, the~evolution of the qubit plus environment ($\vert \cdots \rangle_q
\vert \cdots
\rangle_e$) is given by a unitary transformation $\hat{T}$, which acts
differently if the qubit is in the ground or excited state,
according to the following rules: $\hat{T}(\vert 0 \rangle_q \vert 0
\rangle_e)=\sqrt{1-p}\vert 0 \rangle_q \vert 0
\rangle_e+\sqrt{p}\vert 0 \rangle_q \vert 1 \rangle_e $ and
$\hat{T}(\vert 1 \rangle_q \vert 0 \rangle_e)=\sqrt{1-p}\vert 0
\rangle_q \vert 0 \rangle_e+\sqrt{p}\vert 0 \rangle_q \vert 2
\rangle_e$ with $p$ being a probability, i.e.,~the environment
subsystem goes to a superposition of the states ($\vert 0
\rangle_e$, $\vert 1 \rangle_e$), or~to ($\vert 0 \rangle_e$, $\vert
2\rangle_e$), if~the environment is in $\vert 0 \rangle_e$, or~$\vert 1 \rangle_e$, respectively~\cite{chuang,caruso}. This two-qubit unitary transformations
result in a nonunitary change when the partial trace over the
environment subsystem is taken:
\[
\left( \begin{array}{cc} 1-\rho_{22} & \rho_{12} \\ \rho_{12}^* & \rho_{22}\end{array}\right) \rightarrow
\left( \begin{array}{cc} 1-\rho_{22} & \rho_{12}(1-p) \\ \rho_{12}^*(1-p) & \rho_{22}\end{array}\right) \, .
\]

When the map is applied a very large number of times ($\rightarrow
\infty$), it is straightforward that the initial state tends to the completely decoherent
state:
\[
\left( \begin{array}{cc} 1-\rho_{22} & \rho_{12} \\ \rho_{12}^* & \rho_{22}\end{array}\right) \rightarrow
\left( \begin{array}{cc} 1-\rho_{22} & 0 \\ 0 & \rho_{22}\end{array}\right) \, ,
\]
with an exponential~convergence.

The other example is the spontaneous-emission (also called the
amplitude-damping) quantum channel. In~this channel, the~dynamics of
the qubit system plus the environment is determined by a unitary
transform $\hat{T}$, which only acts if the qubit system is in the
excited state $\vert 1 \rangle_q$, according to the following rules:
$\hat{T}(\vert 0\rangle_q \vert 0 \rangle_e)=\vert 0\rangle_q
\vert 0 \rangle_e$ and $\hat{T}(\vert 1\rangle_q \vert 0
\rangle_e)=\sqrt{1-p}\vert 1\rangle_q \vert 0
\rangle_e+\sqrt{p}\vert 0\rangle_q \vert 1 \rangle_e$, where $p$ is
the probability~\cite{chuang,caruso}. This channel then defines a nonunitary evolution
over the qubit subsystem, which transforms the qubit density matrix
as follows:
\[
\left( \begin{array}{cc} 1-\rho_{22} & \rho_{12} \\ \rho_{12}^* & \rho_{22}\end{array}\right) \rightarrow
\left( \begin{array}{cc} 1-(1-p)\rho_{22} & \rho_{12}\sqrt{1-p} \\ \rho_{12}^*\sqrt{1-p} & (1-p)\rho_{22}\end{array}\right) \, .
\]

If this channel is applied a very large number of times $(\rightarrow \infty)$, the~density matrix converges to a ground state, i.e.,
\[
\left( \begin{array}{cc} 1-\rho_{22} & \rho_{12} \\ \rho_{12}^* & \rho_{22}\end{array}\right) \rightarrow
\left( \begin{array}{cc} 1 & 0 \\ 0 & 0\end{array}\right) \, .
\]

In addition to these examples, there exists another type of quantum
channel defined in the theory of interaction between a
quantum system and an environment, which can be considered~\cite{chuang,caruso}.

It is possible to demonstrate that phase damping and spontaneous-emission quantum channels for qubits
$\hat{\rho}_1, \ldots, \hat{\rho}_6$ in Equation~(\ref{qubits}) can be
obtained by the use of particular unitary transformations of a qutrit
state $\hat{\rho}$. To~justify this, we assumed a two-qubit quantum
system where one of the levels cannot be populated, i.e.,~the
4 $\times$ 4 density matrix has an eigenvalue equal to zero, e.g.,

\begin{equation}
\hat{\rho}=\left(
\begin{array}{cccc}
\rho_{11} & \rho_{12} & \rho_{13} & 0 \\
\rho_{21} & \rho_{22} & 0 & 0 \\
\rho_{31} & 0 & \rho_{33} & 0 \\
0 & 0 & 0 & 0
\end{array}
 \right) ;
 \label{sep1}
\end{equation}
it is clear that this density matrix is separable since
$\rho_{23}=\rho_{32}^*=0$. The~partial density matrices can be
operated locally by unitary transformations of the form $\hat{u}_1
\otimes \hat{u}_2$. When only one of the qubits is operated, i.e.,
when the unitary matrix corresponds to a controlled
operation~\cite{chuang}: $\hat{u}_1=\hat{I}$ or $\hat{u}_2=\hat{I}$.
If~$\hat{u}_2=\hat{I}$, then the unitary transformation only operates
over the second qubit,
\begin{equation}
\hat{u}=\left(\begin{array}{cccc}
u_{11} & u_{12} & 0 & 0 \\
u_{21} & u_{22} & 0 & 0 \\
0 & 0 & u_{11} & u_{12} \\
 0 & 0 & u_{21} & u_{22}
\end{array}\right) .
\label{uni1}
\end{equation}

By means of this type of unitary matrix, one can define an operation
in the qutrit system that approximately only affects $\hat{\rho}_2$.
This is done by ignoring the fourth row and the fourth column of
(\ref{sep1}); the~resulting qutrit state is then operated by the
unitary matrix resulting from the elimination of the fourth row and
the fourth column of Equation~(\ref{uni1}). For~the operator to be still
unitary, the $(3,3)$ entry must be replaced by one. Following these
and other analogous arguments, we study the application of the
unitary transforms:
\begin{equation}
\hat{U}_1=\left( \begin{array}{ccc} u_{11} & u_{12} & 0 \\ u_{21} & u_{22} & 0 \\ 0 & 0 & 1 \end{array} \right) , \quad
\hat{U}_2=\left( \begin{array}{ccc} u_{11} & 0 & u_{12} \\ 0 & 1 & 0 \\ u_{21} & 0 &u_{22} \end{array} \right) , \quad
\hat{U}_3=\left( \begin{array}{ccc} 1 & 0 & 0 \\ 0 & u_{11} & u_{12} \\ 0 & u_{21} & u_{22} \end{array} \right) \quad
\label{unit}
\end{equation}
on the qutrit density matrices:
\begin{equation}
\hat{\sigma}_1=\left(\begin{array}{ccc} \rho_{11} & \rho_{12} & \rho_{13} \\ \rho_{21} & \rho_{22} & 0 \\ \rho_{31} & 0 & \rho_{33} \end{array} \right) , \quad
\hat{\sigma}_2=\left(\begin{array}{ccc} \rho_{11} & 0 & \rho_{13} \\ 0 & \rho_{22} & \rho_{23} \\ \rho_{31} & \rho_{32} & \rho_{33} \end{array} \right) , \quad
\hat{\sigma}_3=\left(\begin{array}{ccc} \rho_{11} & \rho_{12} & 0 \\ \rho_{21} & \rho_{22} & \rho_{23} \\ 0 & \rho_{23} & \rho_{33} \end{array} \right) \, .
\label{dens}
\end{equation}

{ The unitary transformations in Equation~(\ref{unit}) can be enacted on any of the density matrices in Equation~(\ref{dens}), which define a nonunitary transformation of the qubits defined in Equation~(\ref{qubits}). These~qubit transformations are found by the substitution of Equations~(\ref{unit}) and (\ref{dens}) into Equation~(\ref{primes}), e.g.,~the unitary transformation $\hat{U}_1^\dagger \hat{\sigma}_1 \hat{U}_1$ results in the following transformations of the qubits:
\[
\begin{array}{ll}
&\hat{\rho}_1^\prime=\left(\begin{array}{cc}1-\rho_{33} & \rho_{13} \, u_{11}^*\\ \rho_{31} \, u_{11} & \rho_{33}\end{array}\right) , \nonumber \\
&\hat{\rho}_2^\prime=\left(\begin{array}{cc}1 - u_{12}^* (\hat{\sigma}_1 \hat{U}_1)_{12} - u_{22}^* (\hat{\sigma}_1 \hat{U}_1)_{22} & u_{11}^* (\hat{\sigma}_1 \hat{U}_1)_{12} + u_{21}^* (\hat{\sigma}_1 \hat{U}_1)_{22} \\ u_{12}^* (\hat{\sigma}_1 \hat{U}_1)_{11} + u_{22}^* (\hat{\sigma}_1 \hat{U}_1)_{21} & u_{12}^* (\hat{\sigma}_1 \hat{U}_1)_{12} + u_{22}^* (\hat{\sigma}_1 \hat{U}_1)_{22} \end{array}\right) , \nonumber \\
&\hat{\rho}_3^\prime = \left(\begin{array}{cc} u_{11}^* (\hat{\sigma}_1 \hat{U}_1)_{11} + u_{21}^* (\hat{\sigma}_1 \hat{U}_1)_{21} & \rho_{13} \, u_{11}^*\\ \rho_{31} \, u_{11} & 1 - u_{11}^* (\hat{\sigma}_1 \hat{U}_1)_{11} - u_{21}^* (\hat{\sigma}_1 \hat{U}_1)_{21} \end{array}\right) , \nonumber \\
&\hat{\rho}_4^\prime = \left( \begin{array}{cc}
u_{12}^* (\hat{\sigma}_1 \hat{U}_1)_{12} + u_{22}^* (\hat{\sigma}_1 \hat{U}_1)_{22} & \rho_{13} u_{12}^* \\
\rho_{31} u_{12} & 1-u_{12}^* (\hat{\sigma}_1 \hat{U}_1)_{12} - u_{22}^* (\hat{\sigma}_1 \hat{U}_1)_{22}
\end{array} \right) \, ,\nonumber \\
&\hat{\rho}_5^\prime =\left( \begin{array}{cc}
u_{11}^* (\hat{\sigma}_1\hat{U}_1)_{11} + u_{21}^* (\hat{\sigma}_1\hat{U}_1)_{21} & u_{11}^* (\hat{\sigma}_1\hat{U}_1)_{12} + u_{21}^* (\hat{\sigma}_1\hat{U}_1)_{22} \\
u_{12}^* (\hat{\sigma}_1\hat{U}_1)_{11} (r11 u11 + r12 u21) + u_{22}^* (\hat{\sigma}_1\hat{U}_1)_{21} & 1-u_{11}^* (\hat{\sigma}_1\hat{U}_1)_{11} - u_{21}^* (\hat{\sigma}_1\hat{U}_1)_{21}
\end{array}
\right) , \\
&\hat{\rho}_6^\prime =\left(\begin{array}{cc}
1-\rho_{33} & \rho_{13} \, u_{12}^* \\ \rho_{31} \, u_{12} & \rho_{33}
\end{array}
\right).
\end{array}
\]

From these results, one can notice that the transformed qubits $\hat{\rho}_1^\prime$ and $\hat{\rho}_6^\prime$ correspond to the phase damping channel of $\hat{\rho}_1$ with different damping parameters. Furthermore, the~qubit states $\hat{\rho}_2^\prime$, $\hat{\rho}_5^\prime$ can be seen as quasi-unitary transformations of the initial states $\hat{\rho}_2$, $\hat{\rho}_5$, respectively. In~a similar way, one can obtain all the possible unitary transformations of the density matrices in Equation~(\ref{dens}). These~transformations lead to the identification of two types of quantum channels: the phase damping and a quasi-unitary operation described below.
}

The unitary transformation over the density matrices $\hat{\sigma}_1$,
$\hat{\sigma}_2$, and~$\hat{\sigma}_3$ results in a change over their associated
qubits $\hat{\rho}_1, \ldots, \hat{\rho}_6$, to~$\hat{\rho}_1^\prime, \ldots, \hat{\rho}_6^\prime$, which denote the
qubits after the transformation. We have found the following interesting expressions:
\begin{eqnarray}
\hat{U}_1^\dagger \hat{\sigma_1} \hat{U}_1 \Rightarrow \hat{\rho}_1^\prime =\left( \begin{array}{cc}1-\rho_{33} & u_{11}^* \rho_{13} \\
u_{11} \rho_{31} & \rho_{33} \end{array}\right), \quad \hat{\rho}_6^\prime =\left( \begin{array}{cc}1-\rho_{33} & u_{12}^* \rho_{13} \\
u_{12} \rho_{31} & \rho_{33} \end{array}\right); \nonumber \\
\hat{U}_2^\dagger \hat{\sigma_1} \hat{U}_2 \Rightarrow \hat{\rho}_2^\prime =\left( \begin{array}{cc}1-\rho_{22} & u_{11}^* \rho_{12} \\
u_{11} \rho_{21} & \rho_{22} \end{array}\right), \quad \hat{\rho}_4^\prime =\left( \begin{array}{cc}1-\rho_{33} & u_{12}^* \rho_{12} \\
u_{12} \rho_{21} & \rho_{33} \end{array}\right); \nonumber \\
\hat{U}_2^\dagger \hat{\sigma_2} \hat{U}_2 \Rightarrow \hat{\rho}_2^\prime =\left( \begin{array}{cc}1-\rho_{22} & u_{21}^* \rho_{32} \\
u_{21} \rho_{23} & \rho_{22} \end{array}\right), \quad \hat{\rho}_4^\prime =\left( \begin{array}{cc}\rho_{22} & u_{22} \rho_{23} \\
u_{22}^* \rho_{32} & 1-\rho_{22} \end{array}\right); \\
\hat{U}_3^\dagger \hat{\sigma_2} \hat{U}_3 \Rightarrow \hat{\rho}_3^\prime =\left( \begin{array}{cc}\rho_{11} & u_{22} \rho_{13} \\
u_{22}^* \rho_{31} & 1-\rho_{11} \end{array}\right), \quad \hat{\rho}_5^\prime =\left( \begin{array}{cc}\rho_{11} & u_{21} \rho_{13} \\
u_{21}^* \rho_{31} & 1-\rho_{11} \end{array}\right); \nonumber \\
\hat{U}_1^\dagger \hat{\sigma_3} \hat{U}_1 \Rightarrow \hat{\rho}_1^\prime =\left( \begin{array}{cc}1-\rho_{33} & u_{21}^* \rho_{23} \\
u_{21} \rho_{32} & \rho_{33} \end{array}\right), \quad \hat{\rho}_6^\prime =\left( \begin{array}{cc}1-\rho_{33} & u_{22}^* \rho_{23} \\
u_{22} \rho_{32} & \rho_{33} \end{array}\right); \nonumber \\
\hat{U}_3^\dagger \hat{\sigma_3} \hat{U}_3 \Rightarrow \hat{\rho}_3^\prime =\left( \begin{array}{cc}\rho_{11} & u_{12} \rho_{12} \\
u_{12}^* \rho_{21} & 1-\rho_{11} \end{array}\right), \quad \hat{\rho}_5^\prime =\left( \begin{array}{cc}\rho_{11} & u_{11} \rho_{12} \nonumber \\
u_{11}^* \rho_{21} & 1-\rho_{11} \end{array}\right).
\label{phdm}
\end{eqnarray}

In most of the cases, the~resulting qubits $\hat{\rho}^\prime_j$
correspond to the phase damping quantum channel of $\hat{\rho}_j$, as can be seen in Expression (\ref{phdm}).
In this channel, the~probability amplitudes given by the original
off-diagonal terms of the qubits are multiplied by a number. {The damping parameters are associated with different entries of the unitary transformation $u_{jk}$, which in general are complex numbers. When~the unitary transformation correspond to a real matrix, then the expression for the standard phase damping map is obtained}. As~you can see in Equation~(\ref{phdm}), in some 
cases, the~unitary transformations leads to the quantum channel of another
qubit, e.g.,~after the application of $\hat{U}_1$ to
$\hat{\sigma}_1$, the~qubit $\hat{\rho}_6^\prime$ is the phase
damping channel of $\hat{\rho}_1$. Furthermore, in~some other cases, the~obtained density matrices correspond to transformations similar to the
phase damping channel of matrices outside the ones in
Equation~(\ref{qubits}), e.g.,~$\hat{\rho}_4^\prime$ after the application of $\hat{U}_2$ to $\hat{\sigma}_1$.
Although these matrices seem unrelated, they have the same form as
the phase damping channel. In~the case of $\hat{U}$ being
a rotation matrix with a time-dependent angle $\theta=\omega t$, the~original qubit states can be recovered at the time $t=2 \pi l
/\omega$, $l=0,1,2,\ldots$. 

The unitary transformations ($\hat{U}_1$, $\hat{U}_2$, $\hat{U}_3$)
previously described can also lead to quasi-unitary transformations of
the qubits. In~particular, for the unitary transformation $\hat{U}_1^\dagger \hat{\sigma}_1 \hat{U}_1$, one gets the quasi-unitary transformations: 
{
\begin{eqnarray}
 \hat{\rho}_2^\prime&=&\hat{\mathcal{U}}^{ \dagger} \hat{\rho}_2 \,\hat{\mathcal{U}}+\rho_{33} \left( \begin{array}{cc} \vert u_{12} \vert^2 & -u_{11}^* u_{12} \\ -u_{11} u_{12}^* & -\vert u_{12} \vert^2 \end{array} \right)\, , \nonumber \\
 \hat{\rho}_5^\prime &=& \hat{\mathcal{U}}^\dagger \hat{\rho}_5 \, \hat{\mathcal{U}}+\rho_{33} \left(\begin{array}{cc} -\vert u_{21} \vert^2 & -u_{21}^* u_{22} \\ -u_{21} u_{22}^* & \vert u_{21} \vert^2 \end{array} \right) \, , 
 \end{eqnarray}
 with $\hat{\mathcal{U}}=\left( \begin{array}{cc} u_{11} & u_{12} \\
u_{21} & u_{22} \end{array}\right)$ being a two-dimensional unitary transformation. For~the other qubits, one can also define quasi-unitary transformations as~follows:
\begin{itemize}
\item [(a)] From the qutrit unitary transformation $\hat{U}_1^\dagger \hat{\sigma}_3 \hat{U}_1$,
\begin{eqnarray}
 \hat{\rho}_2^\prime &=& \hat{\mathcal{U}}^\dagger \hat{\rho}_2 \, \hat{\mathcal{U}}+\rho_{33} \left(\begin{array}{cc} -\vert u_{12} \vert^2 & u_{11}^* u_{12} \\ u_{11} u_{12}^* & \vert u_{12} \vert^2 \end{array} \right) \, , \nonumber \\
 \hat{\rho}_5^\prime &=& \hat{\mathcal{U}}^\dagger \hat{\rho}_5 \, \hat{\mathcal{U}}+\rho_{33} \left(\begin{array}{cc} \vert u_{21} \vert^2 & u_{21}^* u_{22} \\ u_{21} u_{22}^* & -\vert u_{21} \vert^2 \end{array} \right) \, , 
 \end{eqnarray}
 \item [(b)] For the transformation $\hat{U}_2^\dagger \hat{\sigma}_1 \hat{U}_2$,
\begin{eqnarray}
 \hat{\rho}_1^\prime&=&\hat{\mathcal{U}}^{ \dagger} \hat{\rho}_1 \,\hat{\mathcal{U}}+\rho_{22} \left( \begin{array}{cc} \vert u_{12} \vert^2 & -u_{11}^* u_{12} \\ -u_{11} u_{12}^* & -\vert u_{12} \vert^2 \end{array} \right)\, , \nonumber \\
\hat{\rho}_3^\prime&=&\hat{\mathcal{U}}^{ \dagger} \hat{\rho}_3 \, \hat{\mathcal{U}}+\rho_{22} \left( \begin{array}{cc} -\vert u_{21} \vert^2 & -u_{21}^* u_{22} \\ -u_{21} u_{22}^* & \vert u_{21} \vert^2 \end{array} \right)\, .
\end{eqnarray}
\item [(c)] For the transformation $\hat{U}_2^\dagger \hat{\sigma}_2 \hat{U}_2$,
\begin{eqnarray}
\hat{\rho}_1^\prime&=&\hat{\mathcal{U}}^{ \dagger} \hat{\rho}_1 \,\hat{\mathcal{U}}+\rho_{22} \left( \begin{array}{cc} \vert u_{12} \vert^2 & -u_{11}^* u_{12} \\ -u_{11} u_{12}^* & -\vert u_{12} \vert^2 \end{array} \right)\, , \nonumber \\
\hat{\rho}_3^\prime&=&\hat{\mathcal{U}}^{ \dagger} \hat{\rho}_3 \,\hat{\mathcal{U}}+\rho_{22} \left( \begin{array}{cc} -\vert u_{21} \vert^2 & -u_{21}^* u_{22} \\ -u_{21} u_{22}^* & \vert u_{21} \vert^2 \end{array} \right)\, ,
\end{eqnarray}
\item [(d)] From $\hat{U}_3^\dagger \hat{\sigma}_2 \hat{U}_3$,
\begin{eqnarray}
\hat{\rho}_4^\prime&=&\hat{\mathcal{U}}^\dagger \hat{\rho}_4\, \hat{\mathcal{U}}+\rho_{11}\left( \begin{array}{cc} -\vert u_{21} \vert^2 & -u_{12}^* u_{22} \\ -u_{21} u_{22}^* & \vert u_{21} \vert^2 \end{array}\right) \, , \nonumber \\
\hat{\rho}_6^\prime&=&\hat{\mathcal{U}}^\dagger \hat{\rho}_6 \,\hat{\mathcal{U}}+\rho_{11}\left( \begin{array}{cc} -\vert u_{12} \vert^2 & u_{11}^* u_{12} \\ u_{11} u_{12}^* & \vert u_{12} \vert^2 \end{array}\right) \, . 
\end{eqnarray}
\item [(e)] Finally, for~$\hat{U}_3^\dagger \hat{\sigma}_3 \hat{U}_3$,
\begin{eqnarray}
 \hat{\rho}_4^\prime&=&\hat{\mathcal{U}}^\dagger \hat{\rho}_4\, \hat{\mathcal{U}}+\rho_{11}\left( \begin{array}{cc} -\vert u_{21} \vert^2 & -u_{12}^* u_{22} \\ -u_{21} u_{22}^* & \vert u_{21} \vert^2 \end{array}\right) \, , \nonumber \\
\hat{\rho}_6^\prime&=&\hat{\mathcal{U}}^\dagger \hat{\rho}_6 \,\hat{\mathcal{U}}+\rho_{11}\left( \begin{array}{cc} -\vert u_{12} \vert^2 & u_{11}^* u_{12} \\ u_{11} u_{12}^* & \vert u_{12} \vert^2 \end{array}\right) \, ,
\end{eqnarray}
\end{itemize}

For all the cases, $\hat{\mathcal{U}}$ is a two-dimensional unitary transformation.}

As in the phase-damping case, one can think of a rotation matrix
with a time-dependent angle $\theta= \omega t$ as the unitary
operation, i.e.,
\[
\hat{\mathcal{U}}=\left(\begin{array}{cc} \cos (\omega t) & -\sin (\omega t) \\ \sin (\omega t) & \cos (\omega t)\end{array} \right) \, ,
\]
 which, in~the case where $ t\approx 0$, results in the following
 transformations:
\begin{equation}
 \hat{\rho}_j^\prime= \hat{\mathcal{U}}^\dagger \hat{\rho}_j \, \hat{\mathcal{U}}- \rho_{kk} \, \omega \, t \,\hat{\sigma}_x+\mathcal{O}(t^2) \, ,
 \label{quasi}
 \end{equation}
where $\hat{\sigma}_x$ is the Pauli matrix and~$\rho_{kk}$ is a
diagonal component of $\hat{\rho}$, which depends on $j$. Its value
is $k=2$ for $j=1,3$, $k=3$ for $j=2,5$, and~$k=1$ for $j=4,6$. It
is necessary to point out that, for~$\hat{\rho}_5^\prime$ associated
with $\hat{U}_1^\dagger\hat{\sigma}_3\hat{U}_1$, we need to replace
$\rho_{33}$ with $-\rho_{33}$ in Equation~(\ref{quasi}).

In the case where the density matrices correspond to states, where
one of the accessible levels is not occupied, i.e.,
\begin{equation}
\hat{\sigma}_4=\left( \begin{array}{ccc} \rho_{11} & \rho_{12} & 0 \\ \rho_{21} & \rho_{22} & 0 \\ 0 & 0 & 0\end{array}\right), \quad
\hat{\sigma}_5=\left( \begin{array}{ccc} \rho_{11} & 0 & \rho_{13} \\ 0 & 0 & 0 \\ \rho_{31} & 0 & \rho_{33}\end{array} \right), \quad
\hat{\sigma}_6=\left(\begin{array}{ccc} 0 & 0 & 0 \\ 0 & \rho_{22} & \rho_{23} \\ 0 & \rho_{32} & \rho_{33}\end{array}\right),
\label{rhos1}
\end{equation}
we obtain the expressions:
\begin{eqnarray}
\hat{U}^\dagger_2 \hat{\sigma}_4 \hat{U}_2 \Rightarrow \hat{\rho}^\prime_5=\left( \begin{array}{cc}\rho_{11} \vert u_{11} \vert^2 & \rho_{12} u^*_{11} \\ \rho_{21} u_{11} & 1-\rho_{11} \vert u_{11} \vert^2 \end{array}\right), \quad \hat{\rho}^\prime_6 = \left( \begin{array}{cc}1-\rho_{11} \vert u_{12} \vert^2 & \rho_{21} u_{12} \\ \rho_{12} u^*_{12} & \rho_{11} \vert u_{12} \vert^2 \end{array}\right), \nonumber \\
\hat{U}_3^\dagger \hat{\sigma}_4 \hat{U}_3 \Rightarrow \hat{\rho}^\prime_1 = \left( \begin{array}{cc}1-\rho_{22} \vert u_{12}\vert^2 & \rho_{12} u_{12} \\ \rho_{21} u^*_{12} & \rho_{22} \vert u_{12} \vert^2 \end{array}\right), \quad \hat{\rho}^\prime_2=\left(\begin{array}{cc}1-\rho_{22} \vert u_{11}\vert^2 & \rho_{12} u_{11} \\ \rho_{21} u^*_{11} & \rho_{22} \vert u_{11}\vert^2 \end{array}\right), \nonumber \\
\hat{U}^\dagger_1 \hat{\sigma}_5 \hat{U}_1 \Rightarrow \hat{\rho}^\prime_3= \left( \begin{array}{cc}\rho_{11} \vert u_{11} \vert^2 & \rho_{13} u^*_{11} \\ \rho_{31} u_{11}& 1-\rho_{11} \vert u_{11} \vert^2 \end{array} \right), \quad \hat{\rho}^\prime_4= \left( \begin{array}{cc} \rho_{11} \vert u_{12}\vert^2 & \rho_{13} u^*_{12} \\ \rho_{31} u_{12} & 1-\rho_{11} \vert u_{12}\vert^2 \end{array} \right), \\
\hat{U}^\dagger_3 \hat{\sigma}_5 \hat{U}_3 \Rightarrow \hat{\rho}^\prime_1= \left( \begin{array}{cc}1-\rho_{33} \vert u_{22}\vert^2 & \rho_{13} u_{22}\\ \rho_{31} u^*_{22}& \rho_{33} \vert u_{22} \vert^2\end{array} \right), \quad \hat{\rho}^\prime_2= \left( \begin{array}{cc} 1-\rho_{33} \vert u_{21}\vert^2 & \rho_{13} u_{21}\\ \rho_{31} u^*_{21} & \rho_{33} \vert u_{21}\vert^2 \end{array} \right), \nonumber \\
\hat{U}^\dagger_1 \hat{\sigma}_6 \hat{U}_1 \Rightarrow \hat{\rho}^\prime_3= \left( \begin{array}{cc}\rho_{22} \vert u_{21} \vert^2 & \rho_{23} u^*_{21} \\ \rho_{32} u_{21} & 1-\rho_{22}\vert u_{21} \vert^2\end{array} \right), \quad
\hat{\rho}^\prime_4= \left( \begin{array}{cc} \rho_{22} \vert u_{22}\vert^2 & \rho_{23} u^*_{22}\\ \rho_{32} u_{22} & 1-\rho_{22} \vert u_{22}\vert^2 \end{array} \right), \nonumber \\
\hat{U}^\dagger_2 \hat{\sigma}_6 \hat{U}_2 \Rightarrow \hat{\rho}^\prime_5= \left( \begin{array}{cc} \rho_{33} \vert u_{21}\vert^2 & \rho_{32} u^*_{21}\\ \rho_{23} u_{21} & 1-\rho_{33} \vert u_{21} \vert^2\end{array} \right), \quad
\hat{\rho}^\prime_6= \left( \begin{array}{cc} 1-\rho_{33} \vert u_{22} \vert^2 & \rho_{23} u_{22}\\ \rho_{32} u^*_{22} & \rho_{33} \vert u_{22} \vert^2 \end{array}
\right) \, \nonumber .
\label{spon}
\end{eqnarray}

\begin{figure}[h]
\centering
\includegraphics[scale=0.27]{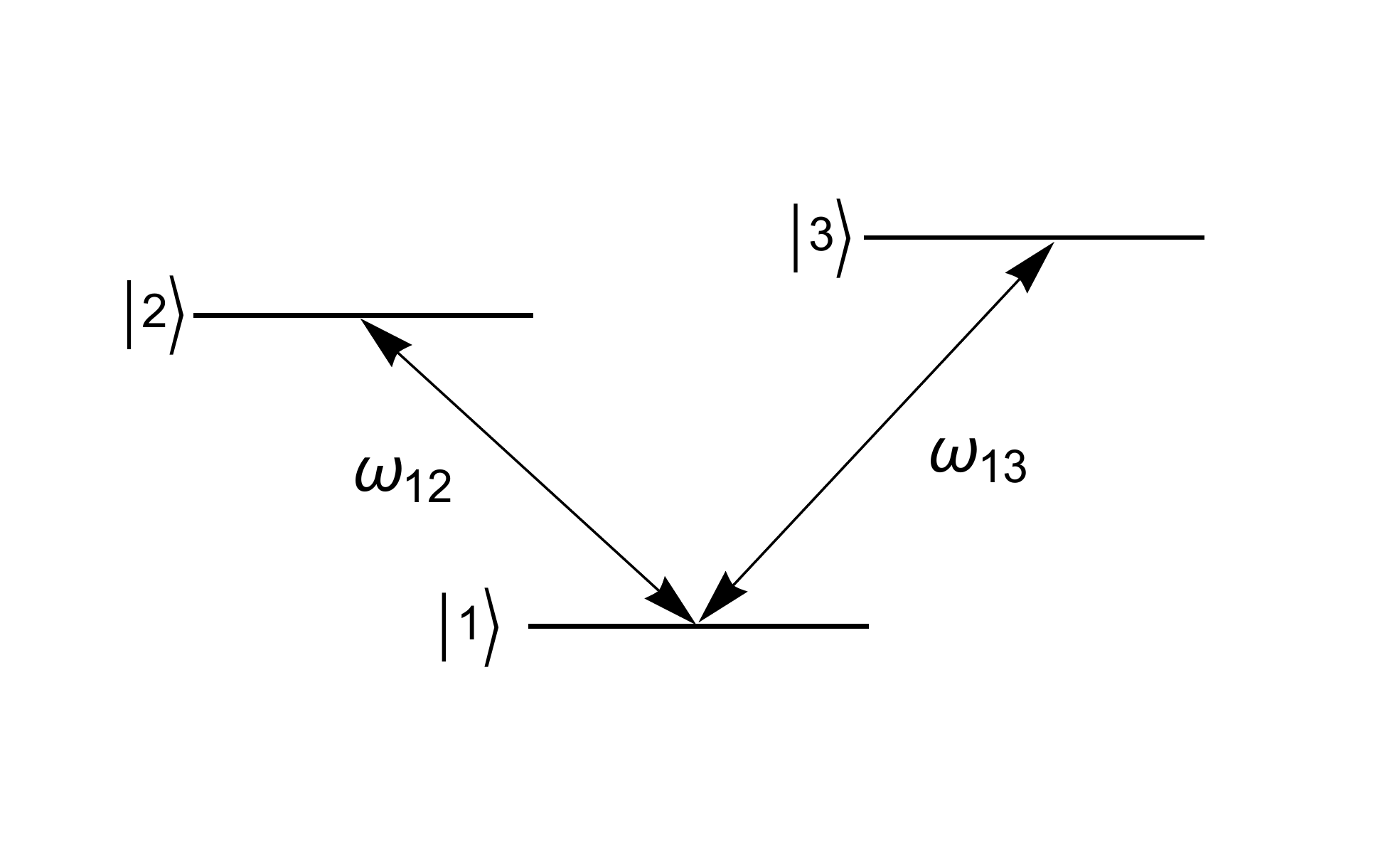}
\includegraphics[scale=0.26]{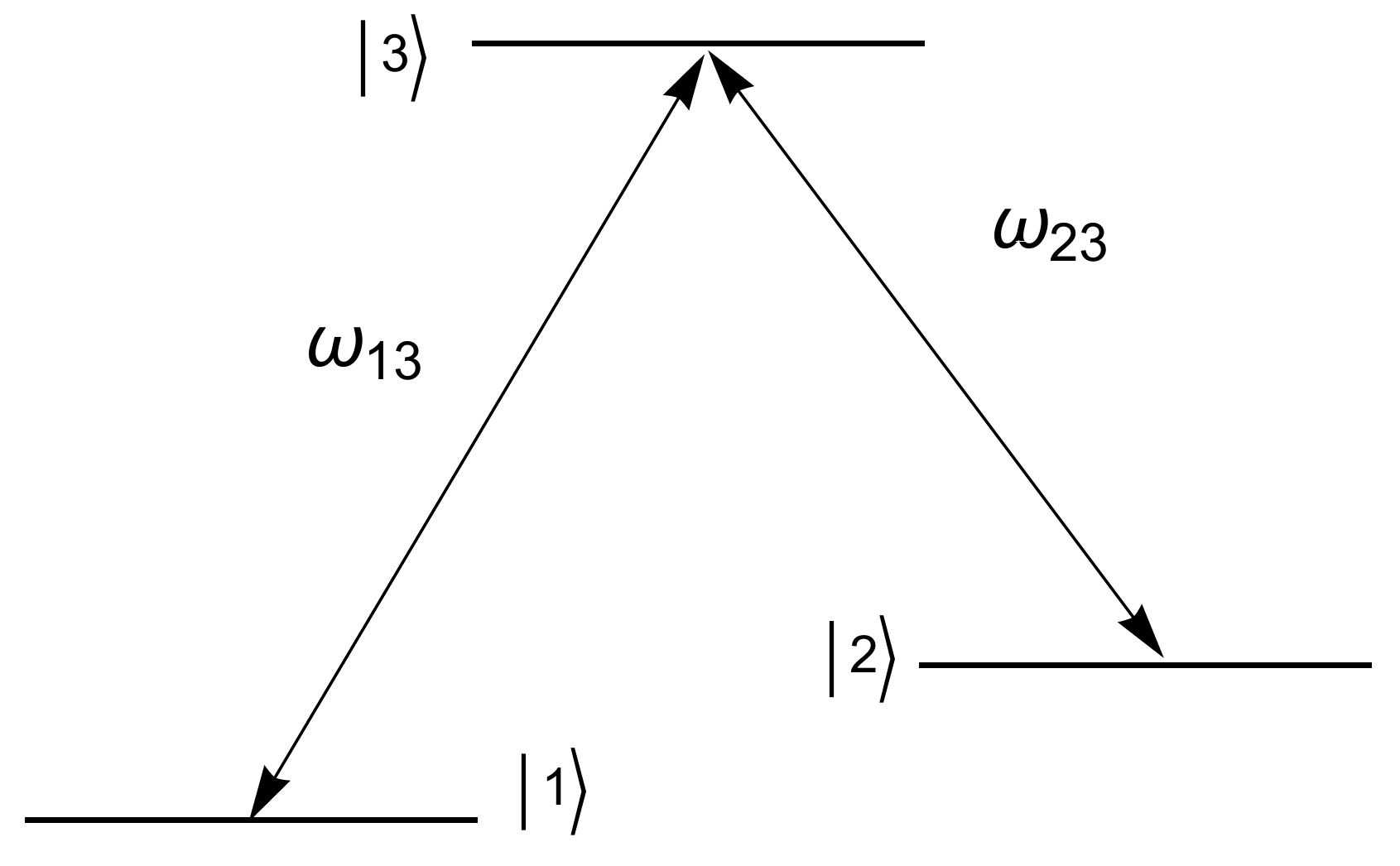}
\includegraphics[scale=0.26]{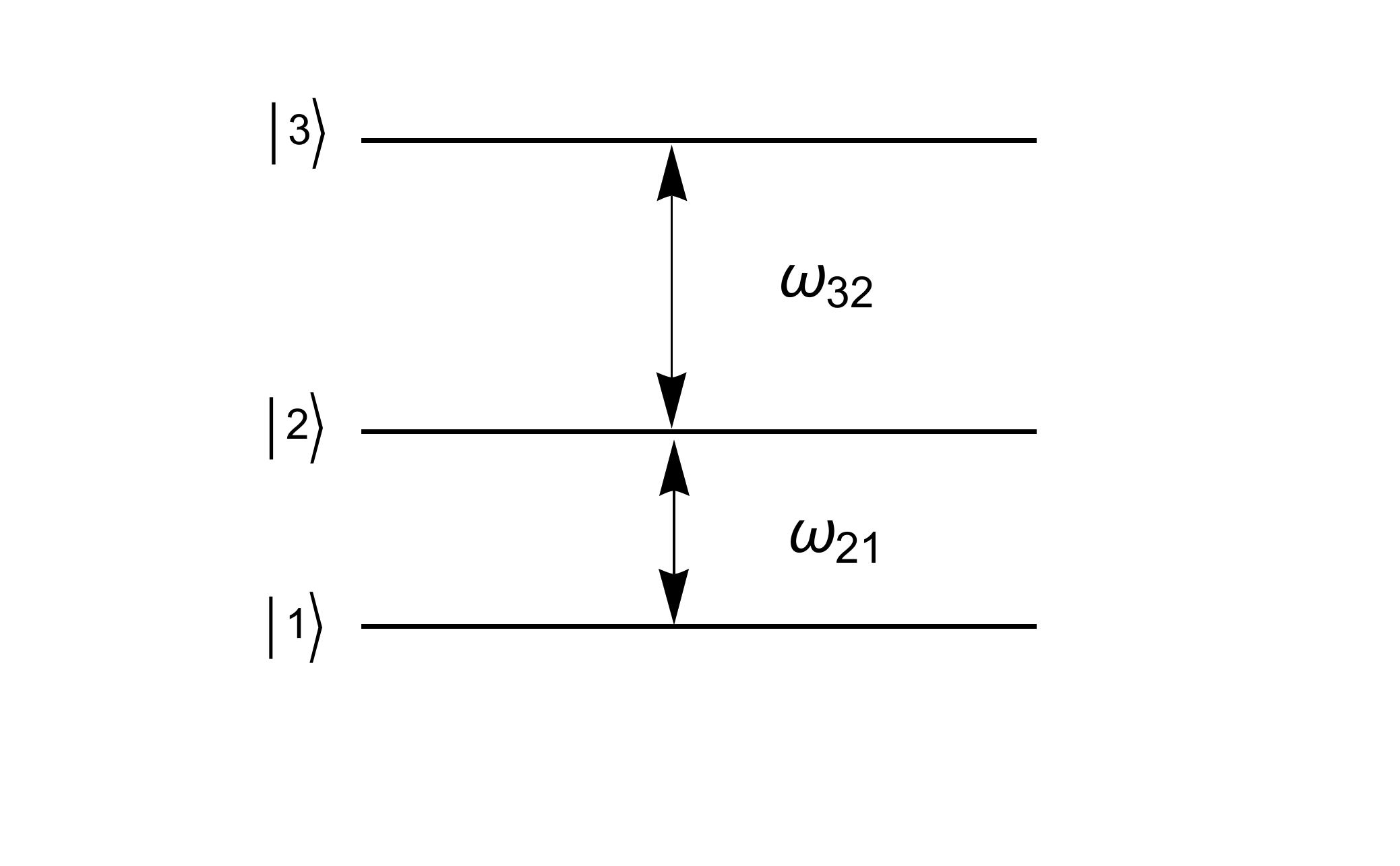}
\caption{State configurations for the V- ({left}), the~$\Lambda$- ({center}), and~the $\Xi$-level ({right})~systems. \label{conf}}
\end{figure}

These transformations in many of the cases can represent the
spontaneous-emission quantum channel. As~in the other examples
studied above, when the unitary matrices are rotated by angle
$\theta=\omega t$, the~original qubit systems can be recovered at
times $t=2 \pi l/\omega$; $l=0,1,2,\ldots$. It is important to
mention that the states represented by Equation~(\ref{rhos1}) correspond
to three-level systems, where~one of the levels is a dark state, and then only two
of the levels can be populated, which have been experimentally
obtained~\cite{exper1976}. These kinds of systems have been of
relevance as they can be created by two-photon processes in a
three-level system~\cite{brewer} or~by the adiabatic variation of the
Rabi frequencies associated with the transitions between the three
states~\cite{zanon}. For~example, to~obtain the state
$\hat{\sigma}_4$, one can think of an atomic $\Lambda$-type
three-level system ($\vert 1 \rangle$, $\vert 2 \rangle$, $\vert 3
\rangle$), which interacts with an environment~\cite{zanon}; see
Figure~\ref{conf}. The~Hamiltonian associated with this system can be written in the form:
\[
\hat{H}=\left(
\begin{array}{ccc}
\omega_1 & 0 & \omega_{13} \\
0 & \omega_2 & \omega_{23} \\
\omega_{13} & \omega_{23} & 0
\end{array}
\right) \, ,
\]
where $\omega_{1,2}$ are the energies of the states $\vert 1\rangle, \vert 2
\rangle$, respectively. By~considering the energy of the ground state $\vert 3 \rangle$ equal to zero,~$\omega_{13}$ and $\omega_{23}$ are the transition energies. Taking
the zero energy in the ground state $\vert 3 \rangle $, we can make
the replacements $\omega_{13} \rightarrow \omega_1 \, e^{-i
\omega_1 t}$ and $\omega_{23}\rightarrow \omega_2\,e^{-i \omega_2 t}$.
The time evolution of the density matrix can be obtained by the
expression:
\begin{equation}
\frac{d}{dt} \hat{\rho}= i [ \hat{\rho}, \hat{H} ]+ \hat{\rho}^\prime \, ,
\label{diff}
\end{equation}
where the matrix $\hat{\rho}^\prime$ is given by the interaction of
the original density matrix with the environment:
\[
\hat{\rho}^\prime= \left(
\begin{array}{ccc}
\gamma_{31} \rho_{33} & -\gamma^\prime \rho_{12} & -\gamma_1 \rho_{13} \\
-\gamma^\prime \rho_{21} & \gamma_{32} \rho_{33} & -\gamma_2 \rho_{23} \\
-\gamma_1 \rho_{31} & -\gamma_2 \rho_{32} & -\gamma \rho_{33}
\end{array}
\right),
\]
where the parameters $\gamma_{31}$, $\gamma_{32}$, and~$\gamma$ are
the spontaneous-emission rates, which must satisfy
\mbox{$\gamma=\gamma_{31}+\gamma_{32}$}, and~the relaxation terms for the
coherence components are named $\gamma_1$ and $\gamma_2$, which also
satisfy $\gamma^\prime=\gamma_1+\gamma_2$. The~resulting
differential equations~(\ref{diff}) can be reduced by considering that the
variation of the parameters $\rho_{13}$, $\rho_{23}$, and~$\rho_{33}$ over time is smaller compared to the spontaneous
emission and decoherence terms $\gamma_{31}$ and $\gamma_{32}$; this
is called the adiabatic hypothesis. Under~this hypothesis, it~is
possible to obtain a state with $\rho_{13}=\rho_{23}=\rho_{33}=0$,
as the solution of the evolution of the density matrix
$\hat{\sigma}_4$ discussed~above. 

Another way to obtain these types
of systems is the case where the environmental interaction is neglected, i.e.,~$\hat{\rho}^\prime=0$ in Equation~(\ref{diff}). The~corresponding Schr\"odinger equation is \mbox{$i \,\frac{d \vert \psi
\rangle}{dt}=\hat{H}\vert \psi\rangle$}, with~$\vert \psi \rangle= a_1 (t) e^{-i \omega_1 t} \vert
1 \rangle+ a_2 (t) e^{-i \omega_2 t} \vert 2 \rangle + a_3 (t)
\vert 3 \rangle$, which in view of the initial conditions
\mbox{$a_1(0)=\frac{\omega_2}{\sqrt{\omega_1^2+\omega_2^2}}$},
$a_2(0)=-\frac{\omega_1}{\sqrt{\omega_1^2+\omega_2^2}}$, $a_3(0)=0$
leads to the solution:
\[
a_1(t)=\frac{\omega_2}{\sqrt{\omega_1^2+\omega_2^2}} \, , \quad
a_2(t)=-\frac{\omega_1}{\sqrt{\omega_1^2+\omega_2^2}} \, ; \quad
a_3 (t)=0 \, ,
\]
so the level $\vert 3 \rangle$ is never~populated. 

The density
matrices $\hat{\sigma}_5$ and $\hat{\sigma}_6$ can be obtained by means of
analogous procedures applied to the V and $\Xi$
configurations of the three-level system depicted in
Figure~\ref{conf}.
\vspace{-8pt}

It is also important to mention that the unitary transformations defined
by the matrices $\hat{U}_1$, $\hat{U}_2$, and~$\hat{U}_3$ in
Equation~(\ref{unit}) can be generated experimentally by different
proposed mechanisms, such as sliding mode control~\cite{SMC09},
adiabatic passage~\cite{AD09,AD11,AD11-1}, and~the robust control
scheme~\cite{RC14,RC17}. {We~want to emphasize that the resulting quasi-unitary evolutions and the different quantum channels obtained in our work can have applications in quantum computing and quantum information theories. We think so because the quasi-unitary operations discussed here could be used as approximations to the standard quantum gates, and furthermore, the obtained quantum channels could also be used in the quantum correction algorithms found in the literature.}

\section{Probability Representation of the Qubit-State~Evolution}\label{sec4}
In the quantum tomographic approach of qubit states~\cite{dod,olga}, the~states are identified with tomographic probability distributions. In~the case of the minimal number of probability parameters, the~density
matrix of the qubit (spin-1/2) state reads~\cite{chernegaA}:
\begin{equation}
\hat{\rho}=\left( \begin{array}{cc}
p_3 & p_1-1/2-i (p_2-1/2) \\
p_1-1/2+i (p_2-1/2) & 1-p_3
\end{array}\right) \, , \quad \sum_{j=1}^3 \left(p_j-\frac{1}{4}\right)^2 \leq \frac{1}{4} \, ,
\label{aa1}
\end{equation}
where $0\leq p_k,\leq1$ with $k=1,2,3$ are the probabilities to obtain the
value $+1/2$ in the $x$, $y$, $z$ axis, respectively. Thus, any
qubit state can be identified through the probabilities $p_1$,
$p_2$, and~$p_3$, i.e.,~given~the density operator, one can get the set $\hat{\rho}\leftrightarrow p_1, p_2, p_3$ and vice~versa.~In the case of qubits~(\ref{qubits})~associated with the qutrit
state, the~evolution of the probabilities after the unitary
operation of the qutrit is determined by Equation~(\ref{primes}).~For~example, we have a probabilistic representation corresponding to
$\hat{\rho}_5^\prime$ in the first formula of Equation~(\ref{spon}), i.e.,
\begin{equation}
p_3 \rightarrow p_3 \vert u_{11} \vert^2 , \quad p_1-1/2-i(p_2-1/2) \rightarrow (p_1-1/2-i(p_2-1/2))u_{11}^* \, .
\label{exx}
\end{equation}

The change of probabilities can be characterized by the evolution of
the Tsallis and Shannon entropies. For~example, in~(\ref{exx}), the
unitary matrix parameter $u_{11}$ determines the evolution of the
Shannon entropy related to a {coin} probability distribution
$(p_3,1-p_3)$ (assume that we have two nonideal classical
coins I and II in such a game as coin flipping, coin tossing, or~heads (up, $\oplus)$ or tails (down, $\ominus)$, which is the
practice of throwing a coin in the air and checking which side is
showing when it lands, in~order to choose between two alternatives
$P_k$ or $(1-P_k)$; $k=1,2$). 
This evolution is of the form:
\[
S(\hat{U})=-p_3 \vert u_{11}\vert^2 \ln \left( p_3 \vert u_{11}\vert^2 \right)-(1-p_3 \vert u_{11}\vert^2) \ln \left( 1-p_3 \vert u_{11}\vert^2 \right) \, .
\]

This entropy, as~a function of the unitary evolution applied to the
qutrit state, characterizes some aspects of the open dynamics of
qubits. We point out that, as~for $p_3$, there exist other classical
entropic characteristics associated with the evolution of $p_1$ and
$p_2$ given by Equation~(\ref{exx}).

\section{Concluding~Remarks}
A new mechanism to study the open system evolution of a
noncomposite qudit system was established. As~an example of
the general procedure, we considered a qutrit system. Associated with the qutrit system, one can
define different qubit density matrices, which evolve in an
open-like way when a unitary transformation is enacted on the~qutrit.

{The application of the resulting transformations for the qubits within the qutrit was also discussed. The~quasi-unitary transformations obtained here might be used as an approximation to quantum gates, whereas the quantum channels could be employed in quantum correction protocols.
}

Different types of quantum channels can be observed using the qubit
decomposition of a qutrit system. In~particular, the~phase damping
and the spontaneous-emission channels were obtained using a unitary
transformation acting on specific qutrit density matrices. The~phase
damping channel was obtained when a unitary transformation of the density
matrix with one off-diagonal term equal to zero was performed. A~spontaneous-emission channel can be observed by unitary transformations
acting over a dark state, i.e.,~a three-level state where one of the
levels cannot be~populated.

In addition to these channels, quasi-unitary transformations of the
qubit states can be defined. This was also done by the application of
a unitary matrix to the generic qutrit~state.

The entropy evolution of the tomographic-probability distributions
determined by the system of qubits was~discussed.

We can extend our analysis to other qudit systems without subsystems
since, o~an arbitrary spin-$j$ density matrix and the spin unitary
evolution, one can associate the smaller spin $j'<j$ evolution.

The possible experimental implementation of the procedure was also
addressed, given that there exist several proposed ways to generate
the unitary transformations such as by sliding mode
control~\cite{SMC09}, adiabatic passage~\cite{AD09,AD11,AD11-1}, or~the robust control scheme~\cite{RC14,RC17}.

\vspace{6pt}

\section*{Acknowledments}
This work was partially supported by DGAPA-UNAM (under Project IN101619).



\begin{thebibliography}{999}
{
\bibitem{weiss}
Weiss, U. \emph{Quantum Dissipative Systems}; World Scientific: Singapore, 1993.

\bibitem{kossa}
Kossakowski, A. On quantum statistical mechanics of non-Hamiltonian systems. \emph{Rep. Math. Phys.} {\bf 1972}, \emph{3}, 247--274.

\bibitem{ingarden}
Ingarden, R.S.; Kossakowski, A. On the connection of nonequilibrium information thermodynamics with non-Hamiltonian quantum mechanics of open systems. \emph{Ann. Phys.} {\bf 1975}, {\em 89}, 451--485.

\bibitem{gorini} 
Gorini, V.; Kossakowski, A.; Sudarshan, E.C.G. Completely positive dynamical semigroups of $N$-level systems. \emph{J. Math. Phys.} {\bf 1976}, {\em 17}, 821--825.

\bibitem{lindblad} 
Lindblad, G. On the generators of quantum dynamical semigroups. \emph{Commun. Math. Phys.} {\bf 1976}, {\em 48}, 119--130.
}
\bibitem{chernegaA}
Chernega, V.N.; Man'ko, O.V.; Man'ko, V.I. Triangle Geometry of the Qubit State in the Probability Representation Expressed in Terms of the Triada of Malevich’s Squares. \emph{J. Russ. Laser Res.} {\bf 2017}, {\em 38}, 141--149.

\bibitem{chernegaB}
Chernega, V.N.; Man'ko, O.V.; Man'ko, V.I. Probability Representation of Quantum Observables and Quantum States. \emph{J. Russ. Laser Res.} {\bf 2017}, {\em 38}, 324--333.

\bibitem{chernegaC}
Chernega, V.N.; Man'ko, O.V.; Man'ko, V.I. Triangle Geometry for Qutrit States in the Probability Representation. \emph{J. Russ. Laser Res.} {\bf 2017}, {\em 38}, 416--425.

\bibitem{entropy18}
L\'opez-Sald\'ivar, J.A.; Castaños, O.; Nahmad-Achar, E.; López-Peña , R.; Man'ko, V.I.; Man'ko, M.A. Geometry and Entanglement of Two-Qubit States in
the Quantum Probabilistic Representation. \emph{Entropy} {\bf 2018}, 20, 630.

\bibitem{QIP}
L\'opez-Sald\'ivar, J.A.; Castaños, O.; Man'ko, M.A.; Man'ko, V.I. Qubit representation of qudit states: Correlations and state reconstruction. \emph{Quantum Inf. Process.} {\bf 2019}, {\em 18}, 210.

\bibitem{wallraff}
Devoret, M.H.; Wallraff, A.; Martinis, J.M. Superconducting Qubits:~A Short Review.~\emph{arXiv} \textbf{2004}, arXiv:cond-mat/0411174v1.

\bibitem{superc}
Devoret, M.H.; Schoelkopf, R.J. Superconducting Circuits for Quantum Information:~An Outlook. \emph{Science}~{\bf 2013}, {\em 339}, 1169--1174.

\bibitem{neeley}
Neeley, M.; Ansmann, M.; Bialczak, R.C.; Hofheinz, M.; Lucero, E.; O'Connell, A.D.; Sank, D.; Wang,~H.; Wenner, J.; Cleland, A.N.; et al.~Emulation of a quantum spin with a
superconducting phase qudit. \emph{Science}~{\bf 2009}, {\em 325}, 722--725.

\bibitem{lanyon}
Lanyon, B.P.; Barbieri, M.; Almeida, M.P.; Jennewein, T.; Ralph, T.C.; Resch, K.J.; Pryde, G.J.; O’Brien, J.L.; Gilchrist, A.; White, A.G. Simplifying quantum logic using
higher-dimensional Hilbert spaces. \emph{Nat. Phys.} {\bf 2009}, 5, 134--140.

\bibitem{chuang}
Chuang, I.L.; Nielsen, M.A. \emph{Quantum Computation and Quantum Information}; Cambridge University Press: Cambridge, UK, 2000.

{
\bibitem{terhal}
Terhal, B.M. Quantum Error Correction for Quantum Memories. \emph{Rev. Mod. Phys.} {\bf 2015}, {\em 87}, 307, doi:10.1103/RevModPhys.87.307.
}

\bibitem{dark1976}
Arimondo, E.; Orriols, G. Nonabsorbing atomic coherences by coherent two-photon transitions in a three-level optical pumping. \emph{Lett. Nuovo C.} {\bf 1976}, {\em 17}, 333--338.
\bibitem{reso1987}
Dalibard, J.; Reynaud, S.; Cohen-Tannoudji, C. La cascade radiative de l'atome habillé. In \emph{Interaction of Radiation with Matter}; A Volume in Honour of Adriano Gozzini; Scuola Normale Superiore: Pisa, Italy, 1987.

\bibitem{eit1990}
Harris, S.E.; Field, J.E.; Imamoglu,~A. Nonlinear Optical Processes Using Electromagnetically Induced~Transparency. \emph{Phys. Rev. Lett.} {\bf 1990}, {\em 64}, 1107, doi:10.1103/PhysRevLett.64.1107.

\bibitem{eit1993}
Harris, S.E. Electromagnetically induced transparency with matched pulses. \emph{Phys. Rev. Lett.} {\bf 1993}, {\em 70}, 552, doi:10.1103/PhysRevLett.70.552.

\bibitem{eit2005}
Fleischhauer, M.; Imamoglu, A.; Marangos, J.P. Electromagnetically induced transparency:~Optics in coherent~media. \emph{Rev. Mod. Phys.} {\bf 2005}, {\em 77}, 633.

{
\bibitem{horo}
Horodecki, M.; Horodecki, P.; Horodecki, R. Separability of mixed states: Necessary and sufficient conditions. \emph{Phys. Lett. A} \textbf{1996}, \emph{223}, 1--8, doi:10.1016/S0375-9601(96)00706-2.

\bibitem{caruso}
Caruso, F.; Giovannetti, V.; Lupo, C.; Mancini, S. Quantum channels and memory effects. \emph{Rev. Mod. Phys.} {\bf 2014}, {\em 86}, 1203.
}


\bibitem{exper1976}
Alzetta, G.; Gozzini, A.; Moi, L.; Orriols, G. An Experimental Method for the Observation of R.F. Transitions and Laser Beat Resonances in Oriented Na Vapour. \emph{Il Nuovo C. B} {\bf 1976}, \emph{36}, 5--20.

\bibitem{brewer}
Brewer, R.G.; Hahn, E.L. Coherent two-photon processes: Transient and steady-state cases. \emph{Phys. Rev. A} {\bf 1975}, \emph{11}, 1641.

\bibitem{zanon}
Zanon-Willette, T.; de Clercq, E.; Arimondo, E. Ultrahigh-resolution spectroscopy with atomic or molecular dark resonances: Exact steady-state line shapes and asymptotic profiles in the adiabatic pulsed regime. \emph{Phys. Rev. A} {\bf 2011}, {\em 84}, 062502, doi:10.1103/PhysRevA.84.062502.

\bibitem{SMC09}
Dong, D.; Petersen, I.R. Sliding mode control of quantum systems. \emph{New J. Phys.} {\bf 2009}, {\em 11}, 105033.

\bibitem{AD09}
Chen, J.-M.; Liang, L.-M.; Li, C.-Z.; Deng, Z.-J. Arbitrary state controlled-unitary gate between two remote atomic qubits via adiabatic passage. \emph{Opt. Commun.} {\bf 2009}, {\em 282}, 4020--4024.

\bibitem{AD11}
Gu\'erin, S.; Hakobyan, V.; Jauslin, H.R. Optimal adiabatic passage by shaped pulses:~Efficiency and~robustness. \emph{Phys. Rev. A} {\bf 2011}, {\em 84}, 013423.

\bibitem{AD11-1}
Torosov, B.T.; Gu\'erin, S.; Vitanov, N.V. High-Fidelity Adiabatic Passage by Composite Sequences of Chirped~Pulses. \emph{Phys. Rev. Lett.} {\bf 2011}, {\em 106}, 233001.

\bibitem{RC14}
Zhang, J.; Greenman, L.; Deng, X.; Whaley, K.B. Robust Control Pulses Design for Electron Shuttling in Solid-State Devices. \emph{IEEE Trans. Control Syst. Technol.} {\bf 2014}, {\em 22}, 2354--2359.

\bibitem{RC17}
Wu, C.; Qi, B.; Chen, C.; Dong, D. Robust Learning Control Design for Quantum Unitary Transformations. \emph{IEEE Trans. Cybern.} {\bf 2017}, {\em 47}, 4405--4417.

\bibitem{dod}
Dodonov, V.V.; Man'ko, V.I. Positive distribution description for spin states. \emph{Phys. Lett. A} {\bf 1997}, {\em 229}, 335--339.

\bibitem{olga}
Man'ko, V.I.; Man'ko, O.V. Spin state tomography. \emph{J. Exp. Theor. Phys.} {\bf 1997}, {\em 85}, 430--434.

\end{thebibliography}
\end{document}